# Giant Positive Magnetoresistance in Co@CoO Nanoparticle Arrays

Hui Xing,[1,2] Wenjie Kong,[1] Chaehyun Kim,[1] Sheng Peng,[3] Shouheng Sun,[3] Zhu-An Xu[2] and Hao Zeng[1,*]

1 Department of Physics, University at Buffalo, the State University of New York, Buffalo, NY 14260

2. Department of Physics, Zhejiang University, Hangzhou, Zhejiang 310027, China

3. Department of Chemistry, Brown University, Providence, RI 02912

We report the magnetotransport properties of self-assembled Co@CoO nanoparticle arrays at temperatures below 100 K. Resistance shows thermally activated behavior that can be fitted by the general expression of $R \propto \exp\{(T_0 / T)^{\nu}\}$. Efros-Shklovskii variable range hopping (ν=1/2) and simple activation (hard gap, ν=1) dominate the high and low temperature region, respectively, with a strongly temperature-dependent transition regime in between. A giant positive magnetoresistance of >1,400% is observed at 10K, which decreases with increasing temperature. The positive MR and most of its features can be explained by the Zeeman splitting of the localized states that suppresses the spin dependent hopping paths in the presence of on-site Coulomb repulsion.

[*] Corresponding author e-mail: haozeng@buffalo.edu

Spin dependent charge transport and magnetoresistance (MR) has been extensively studied due to its technological applications in information industry.[1] Different types of MR, including ordinary MR, giant magnetoresistance (GMR), colossal magnetoresistance, anisotropic magnetoresistance and tunneling magnetoresistance (TMR), have been observed in different materials with a variety of structures, demonstrating rich physics. Of particular interests are granular systems consisting of magnetic grains embedded in a nonmagnetic matrix, which exhibits either GMR with a conducting matrix,[2] or TMR with an insulating matrix,[3] both of which are often negative due to the alignment of the grain magnetization under a magnetic field. Interestingly, anomalous positive MR have been reported in different granular systems.[4,5,6] Possible origins that can account for the positive MR include: ordinary MR caused by the curving of the carrier trajectories in the magnetic field,[7] shrinkage of the wave functions of localized electronic states due to the external field,[8] and suppression of hopping paths due to the Zeeman splitting of the localized states.[9] In a disordered granular system, variable range hopping (VRH) dominates the transport at zero magnetic field, i.e. $R \propto \exp\{(T_0/T)^v\}$. A constant density of states (DOS) near the Fermi level gives Mott VRH[10] with $v = 1/(1+d)$, where d is the dimension of the system. By including the mutual Coulomb interactions between different sites, the DOS near the Fermi level are depleted parabolically, and opening of a soft gap leads to the Efros Shklovskii VRH[11] (ES VRH) with $v = 1/2$. Moreover, opening of a small but finite hard gap around Fermi level caused by the short-range polaron excitations gives $v = 1$.[12] Crossover between these VRH mechanisms is possible when the thermal fluctuation $k_B T$ falls into appropriate energy scales.[13]

Here we report magnetotransport studies in self-assembled Co@CoO nanoparticle arrays as a model granular system. A crossover from ES VRH to simple activation (hard gap) conduction is observed. Giant positive MR with its saturation field increasing with increasing temperature is observed and can be explained by the Zeeman splitting of the localized states that suppresses the spin dependent hopping paths in the presence of a magnetic field.

Monodisperse polycrystalline Co nanoparticles were synthesized by high temperature solution phase method.[14] A typical TEM image of the as-synthesized nanoparticles is shown in Fig.1. As can be seen, each Co particle contains a uniform shell of amorphous cobalt oxide, with approximate stoichiometry of CoO as identified by elemental analysis (not shown here). For charge transport measurements, two lateral electrodes with a gap spacing of 2 µm were patterned by photolithography on Si substrates with a 1 µm thick thermal oxide layer. An appropriate amount of the nanoparticle solutions was deposited between the micro-gap. Induced by controlled evaporation of solvent, the nanoparticles self-assembled into ordered arrays with the interparticle distance of about 2 nm. As-deposited nanoparticle arrays are insulating. After annealing in $N_2$ at 300°C for 1 hour, the surfactants were largely removed, the interparticle distance decreased and the nanoparticle arrays became conducting. MR was measured with magnetic fields applied perpendicular to the substrates.

In the whole temperature range of interest (20-80 K), the resistance shows thermally activated behavior, as can be seen from the exponential dependence of resistance *R* on temperature *T* in Fig. 2(a). *R(T)* can be fitted by the general expression of $R(T) \propto \exp\{(T_0/T)^{\nu}\}$, with different power index extracted for different temperature regions. For the high temperature region (45 K<T<80 K), $\nu$ is found to be ½, which is consistent with ES VRH. This can be seen clearly in Fig. 2(b), where ln(*R)* as a function of $T^{-1/2}$ is fitted by a linear curve. The slope gives the $T_0$ value of 150 K, which corresponds to a Coulomb gap energy $\Delta_C = 13$ meV. It should be noted that the data at this range is slightly noisier than low temperature data and ν is sensitive to the temperature range chosen. ν decreases if we choose a narrower range. At low temperature region (20 K<T< 30K), the linear relationship is obeyed if we choose $\nu = 1.1$, as shown in Fig.2(c). ν ~ 1 indicates the opening of a hard gap at low temperatures. The fitted $T_0$ value is 190 K, which corresponds to the gap energy of $\Delta_H = 16$ meV. Interestingly, in the transition region between 30 and 45 K (Fig. 2(d)), our fitting yields an unexpectedly large power index of $\nu \sim 4$.

Our system can be considered as Co grains embedded in a CoO matrix. Direct tunneling between Co grains, as observed in an earlier work in similar systems,[15] appears to only make a significant contribution at high fields, as discussed later. CoO is known as a Mott insulator.[16] Non-stoichiometric CoO in our system is a disordered semiconductor, where charge transport is dominated by hopping between localized states within the Mott gap. Cross-over from ES VRH to simple activation has been observed earlier in In/$InO_x$ films, where the hardening of the ES gap is due to electron-phonon interactions, and would require that the hard gap $\Delta_H$ be smaller than the soft Coulomb gap $\Delta_C$.[13] In our system, $\Delta_H > \Delta_C$ suggests that different localized states contribute to transport at different temperatures. Upon cooling, some localized states are progressively frozen out, which may contribute to the large index of ν=4 in the transition regime. After the transition, the population of the localized states becomes constant and the system is stabilized with a new gap $\Delta_H$.

Samples show positive MR at temperatures lower than 80 K, which is highly field- and temperature- dependent. Above 80 K, MR becomes vanishingly small. At *T*=10 K, $MR \equiv (R(H) - R(0))/R(0)$ first increases rapidly with increasing field, reaching a maximum value of about 1,400% at H=2.8 Tesla. It then decreases continuously with further increasing field at a slower rate, up to the maximum accessible field of 9 Tesla, as shown in Fig. 3(a). With increasing temperature, the magnitude of MR is significantly reduced, with only 120% remaining at 40 K. Furthermore, the field at which MR shows a maximum increases. At 40 K, no decrease of MR can be observed due to our limited field range. The positive MR is symmetric around the zero field. Knowing that the sample is ferromagnetic at all measuring temperatures, this suggests that the positive MR is not associated with the magnetization of the Co nanoparticles.

Among known mechanisms for positive MR, orbital effect is typically small, particularly for granular systems with small mean free path. Shrinkage of wave function due to applied field has also been used to explain the observed positive MR in disordered systems.[8,13] However, this effect is also small and the field dependence follows $\log(R(H)/R(0)) \sim H^2$ . Another mechanism is Zeeman splitting of the localized states, where spin dependent hopping paths are suppressed by external fields.[9] Briefly speaking, when considering the on-site Coulomb repulsion U, doubly occupied localized states are possible. Matveev *et al*. showed that two kinds of localized sites contribute to the conduction, site A has energy $\varepsilon$ close to the Fermi level $\mu$; Site B has an electron already occupied at a deep level $\varepsilon \sim \mu - U$ , one more electron can occupy the higher energy state close to the Fermi level, forming a spin singlet state.[17] Electron hopping between the two sites is spin and therefore field dependent. At zero field, the localized states accommodate electrons with equal probability of spin up and spin down, and hopping between A and B sites is allowed. An external field alters the energy of A and B sites due to Zeeman splitting, and thus the site occupancy. Both A and B sites are spin polarized and the hopping probability between A and B sites is reduced. With the application of a strong magnetic field, hopping between A and B sites is completely suppressed, leading to increased resistance and therefore positive MR. According to reference [9], $R \propto \exp\left\{ (\frac{T_0}{T})^{1/1+d} f(x) \right\}$, with *f(x)* as a universal function of a single scaling parameter $x = \frac{\mu_B H}{T(T_0/T)^{1/1+d}}$. At low fields, perturbation theory gives $\ln(R(H,T)/R(0,T)) \propto \mu_B H/T$. Indeed, $\ln(R(H,T)/R(0,T))$ at low fields measured at all temperatures are well described by linear dependence on H, as shown in Fig. 3(b). At high fields, deviations from the linear relationship and decrease of resistance are observed, which will be discussed later. However, the linear relationship between $\ln(R(H,T)/R(0,T))$ and $1/T$ is clearly not supported by our data. It should be pointed out that the $H/T$ scaling is derived based on the Mott VRH. It is reasonable that the temperature dependence is invalid in the complex hopping regime in our system, since the DOS is not constant at the Fermi level. Interestingly, all data collapse on a master curve when $\ln(R(H,T)/R(0,T))$ is plotted as a function of $H/T^2$, as shown in Fig. 3(c). This may provide some insight into the temperature dependence of the DOS for the localized sites responsible for the spin dependent hopping.

The giant positive MR up to 1,400% is believed to be due to the s-d exchange coupling between the hopping electrons in the CoO matrix and the magnetic Co core. With the presence of the s-d exchange coupling, the effective g factor increases and thus enhances the Zeeman splitting of the localized states which causes the giant positive MR. The rapid reduction of the magnitude of MR with increasing temperature and the increase of the field value at which MR shows maxima is simply a manifestation of the $H/T^2$ scaling. The decreases of MR at high fields can be explained by the direct spin dependent tunneling between Co nanoparticles. At sufficiently high fields, $f(x)$ attains a constant value,[9] which will give a saturated positive MR. Upon the extensive suppression of the hopping paths in the CoO matrix, direct spin dependent tunneling between Co grains will take over. Its unsaturated nature at high fields can be explained by the non-collinear surface spin effects.[18]

In conclusion, we studied the magnetotransport properties of the self-assembled Co@CoO nanoparticle arrays at temperatures below 100 K. ES VRH and hard gap hopping dominate the high and low temperature regions, respectively. The large power index of $\nu$=4 in the transition region is possibly related to a progressive freezing of some localized states. A giant positive MR of 1,400% is observed and can be interpreted as the Zeeman splitting induced suppression of the spin dependent hopping paths. All MR can be scaled as $H/T^2$. This unusual temperature dependence may be related to non-constant density of states at the Fermi level, and further investigation is needed.

**ACKNOWLEDGMENTS**

This work is supported by NSF grant # DMR-0547036 and University at Buffalo Integrated Nanostructured Systems Instrument Facilities.

**Figure captions:**

Figure 1. A typical TEM image of 12nm Co nanoparticles. Inset shows a magnified image demonstrating the presence of an amorphous surface CoO layer.

Figure 2. (a) Resistance *R* as a function of temperature *T* showing exponential dependence; (b) *ln(R)* as a function of $T^{-0.5}$ at $45\ K<T<80\ K$; (c ) *ln(R)* as a function of $T^{-1.1}$ at $20\ K<T<30\ K$; (d) *ln(R)* as a function of $T^{-4}$ at $30\ K<T<45\ K$; the lines are linear fittings and the top axes are temperature *T* in Fig.s 2(b)-(d).

Figure 3. (a) MR as a function of *H* for T=10 K and 40 K, respectively; (b) *ln(R(H)/R(0))* as a function of *H* for *T* between 10-70 K; the dashed lines are linear fitting in the positive, low field regions; (c) *ln(R(H)/R(0))* as a function of $H/T^2$ measured between 10-80 K; all data collapse onto a master curve.

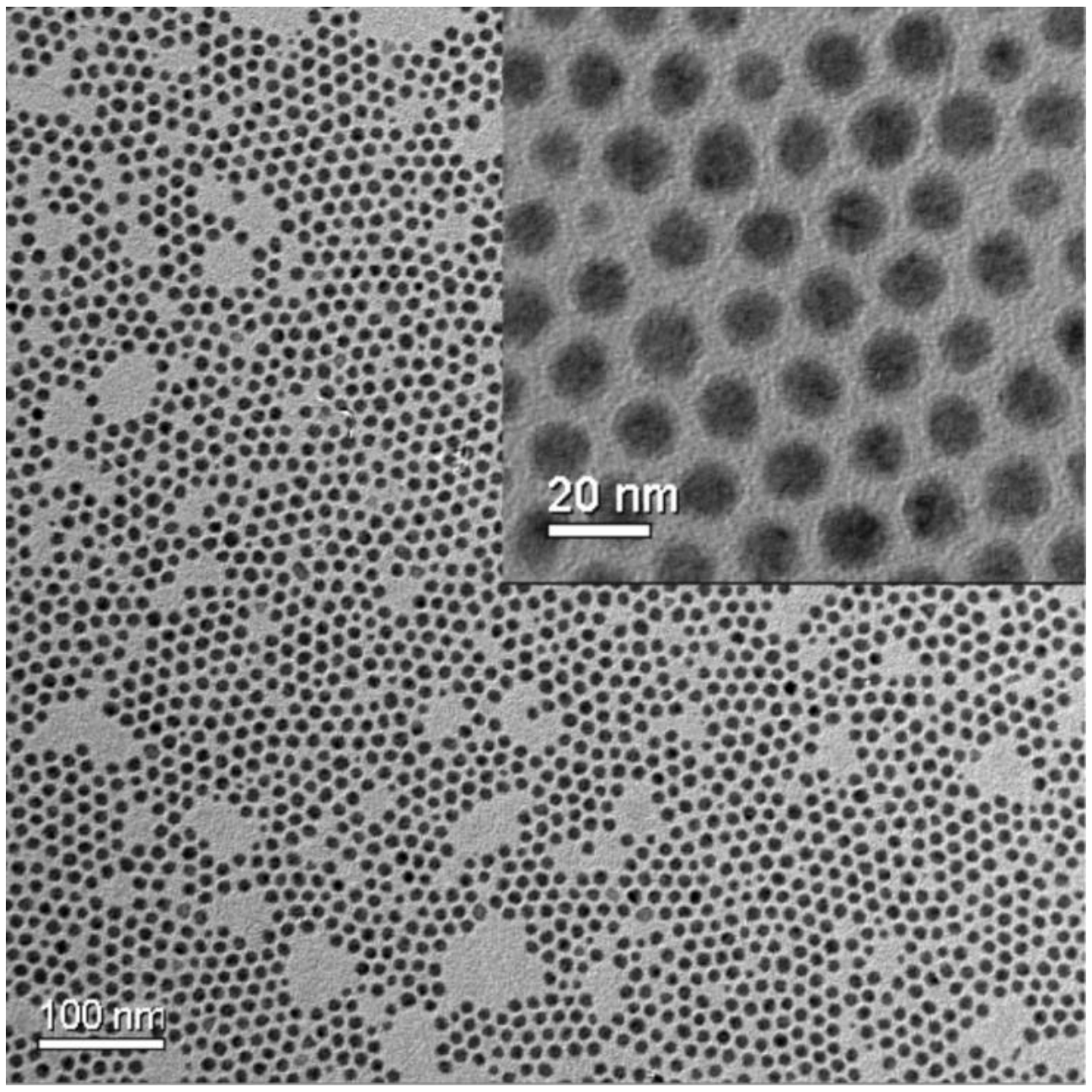

Fig. 1

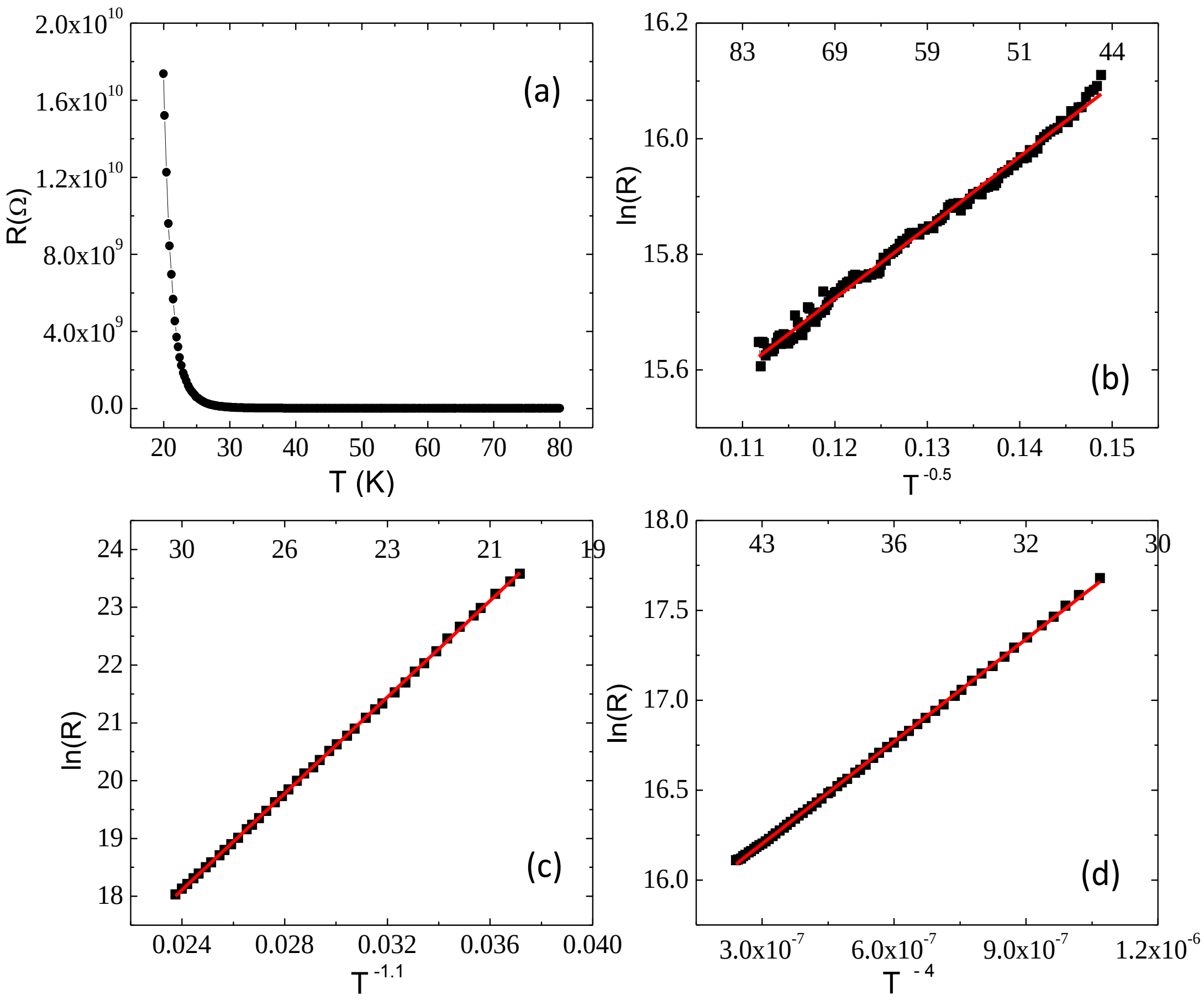

Fig. 2

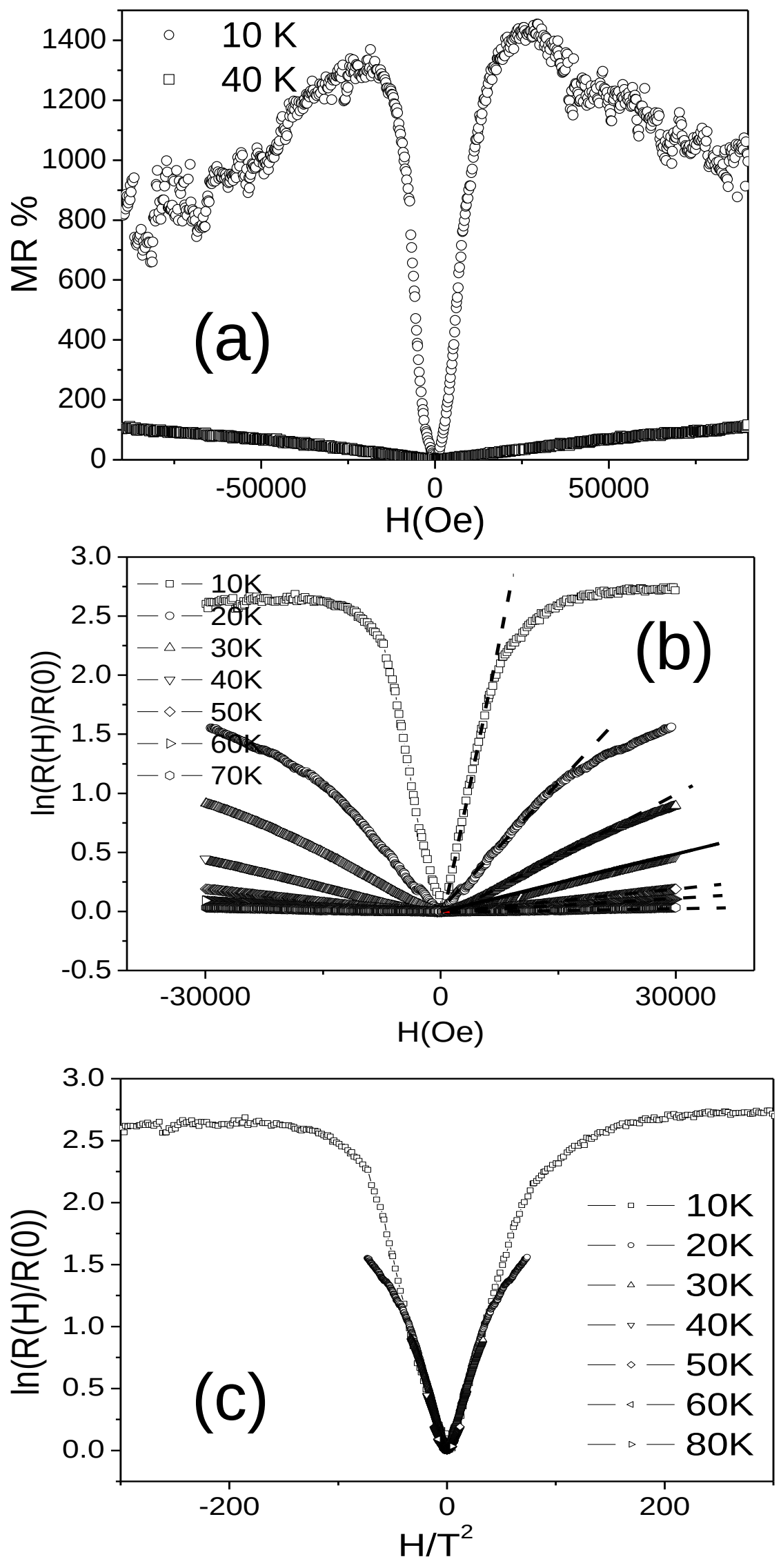

Fig. 3